\documentclass[
    superscriptaddress,
    aip,
	 jap,
    preprintnumbers,
    twocolumn,
    reprint,
    nobalancelastpage,
    nonbibnotes,
    fleqn,
    a4paper,
	 floatfix
]{revtex4-1}
\newcommand{\twocolumnmode}{true}

\usepackage{graphicx}
\usepackage{amsmath}
\usepackage{amsfonts}
\usepackage{color}
\usepackage{ifthen}
\usepackage[
	colorlinks,
	linkcolor=blue,
	anchorcolor=blue,
	citecolor=blue,
	urlcolor=blue,
	menucolor=blue,
	filecolor=blue,
	pagecolor=blue
]{hyperref}

\begin{document}

\preprint{{\em Phys. Rev. B}: Vol. 79, Art. No. 054104, 1 February 2009}

\title{Evidence for dielectric aging due to progressive 180${}^\circ$ domain wall pinning in polydomain Pb(Zr${}_{0.45}$Ti${}_{0.55}$)O${}_3$ thin films}

\author{Pavel \surname{Mokr\'{y}}}
\email{pavel.mokry@tul.cz}%
\affiliation{Institute of Mechatronics and Computer Engineering, Technical University of Liberec, CZ-46117 Liberec, Czech Republic}

\author{Yongli \surname{Wang}}
\altaffiliation[Present address: ]{Ceramic Components Division, EPCOS OHG, Siemensstrasse 43, 8530 Deutschlandsberg, Austria}
\affiliation{Department of Materials, Ceramics Laboratory, EPFL Swiss Federal Institute of Technology, CH-1015 Lausanne, Switzerland}

\author{Alexander K. \surname{Tagantsev}}
\affiliation{Department of Materials, Ceramics Laboratory, EPFL Swiss Federal Institute of Technology, CH-1015 Lausanne, Switzerland}

\author{Dragan \surname{Damjanovic}}
\affiliation{Department of Materials, Ceramics Laboratory, EPFL Swiss Federal Institute of Technology, CH-1015 Lausanne, Switzerland}

\author{Igor  \surname{Stolichnov}}
\affiliation{Department of Materials, Ceramics Laboratory, EPFL Swiss Federal Institute of Technology, CH-1015 Lausanne, Switzerland}

\author{Nava  \surname{Setter}}
\affiliation{Department of Materials, Ceramics Laboratory, EPFL Swiss Federal Institute of Technology, CH-1015 Lausanne, Switzerland}

\begin{abstract}
An evidence that the dielectric aging in the polydomain 
Pb(Zr${}_{0.45}$Ti${}_{0.55}$)O${}_3$ thin films is controlled by progressive 
pinning of 180${}^\circ$ domain walls is presented. To provide such a 
conclusion, we use a general method, which is based on the study of the time 
evolution of the nonlinear, but anhysteretic, dielectric response of the 
ferroelectric to a weak electric field. A thermodynamic model of the 
ferroelectric system where the dielectric response is controlled by bending 
movements of pinned 180${}^\circ$ domain walls is developed. Within this model,
 the nonlinear permittivity of the ferroelectric is expressed as a function of 
the microstructural parameters of the domain pattern. It is shown that by 
using the analysis of the time evolution of the nonlinear permittivity, it is 
possible to estimate changes in the concentration of the pinning centers that 
block the movements of the 180${}^\circ$ domain walls during aging in 
polydomain perovskite ferroelectrics.
\end{abstract}



\date{Received 2 October 2008; revised manuscript received 19 December 2008}

\maketitle


\section{Introduction}
\label{sec_Intro}

Aging of dielectric properties represents an unwanted feature in 
ferroelectric ceramics and lead zirconate-titanate (PZT) thin 
films, in particular, since it prevents their use in devices, 
which require very stable functional properties. It is believed that the 
origin of the dielectric aging in PZT and other similar perovskite 
ferroelectrics is caused by a rearrangement of pinning centers that block the 
movements of domain walls at more points and reduces the domain wall (extrinsic)
 contribution to permittivity of the 
ferroelectric.\cite{Robels.JApplPhys.73.1993} In the following text, we will 
call this concept aging due to \emph{progressive pinning}. Unfortunately, 
examples of credible evidence for the progressive pinning concept are still 
rather scarce
\cite{Robels.JApplPhys.73.1993,Carl.Ferroelectrics.17.1978}
due to strongly limited possibilities of experimental techniques. Even the 
direct observation of domain wall pinning itself is very rare and requires 
quite advanced experimental techniques
\cite{Yang.PhysRevLett.82.1999,Schrade.MechMat.39.2007}.
For this reason, progress in understanding the detailed role of domain wall 
pinning in phenomena such as fatigue, aging, or imprint 
\cite{Warren.JApplPhys.77.1995} is rather complicated, since it relies mainly 
on indirect measurements. One possibility to mention here is 
the study of the domain wall dynamics using measurements of the small-signal 
dielectric response,
\cite{Kleemann.Ferroelectrics.334.2006,Kleemann.AnnuRevMatRes.37.2007,Braun.PhysRevLett.94.2005}
which is very sensitive to the pinning of the domain walls. Unfortunately, 
these experiments were analyzed using models, which cannot serve any 
quantitative information on concentration of the pinning centers at the domain 
wall, etc.

The aforementioned issues have motivated the study presented below, where we 
will apply a general method introduced recently in 
Ref.\cite{Mokry.Ferroelectrics.375.2008} that allows us to provide evidence 
that the dielectric aging of polydomain PZT films is controlled by 
progressive pinning of 180$^{\circ}$ domain walls. The adopted 
method is based on the measurement of the \emph{nonlinear dielectric response} 
of non-polar polydomain ferroelectric samples to a \emph{weak 
electric field}. This approach has several advantages. First, the application 
of a weak electric field to the ferroelectric sample has a minimum effect on 
the aging process. It is known that the electric cycling of the ferroelectric 
sample with electric fields comparable (in magnitude) to the coercive field 
has a strong deaging effect.\cite{Morozov.JApplPhys.2008} Second, the 
dielectric response of the polydomain ferroelectric sample to the weak 
electric field is \emph{anhysteretic},\cite{Li.JApplPhys.69.1991} since it is 
controlled only by a fast reversible movement of the domain walls. This makes 
it possible to adopt a much simpler theoretical treatment in our analysis. 
It follows from the symmetry reasons that in the limit of the 
electric field tending zero, the permittivity $\varepsilon_f$ of the 
polydomain non-polar ferroelectric sample (i.e., with the equal 
volumes of domains with the vectors of spontaneous polarization oriented along 
and against the applied electric field) is quadratically dependent on the 
electric field $E$,
\begin{equation}
  \varepsilon_f(E) = \varepsilon_L + bE^2,
  \label{eq:01:EQuadr}
\end{equation}
where $\varepsilon_L$ is the small-signal permittivity and $b$ is the 
dielectric nonlinearity constant. This is in contrast to the essentially 
hysteretic dielectric response of the ferroelectric to the subswitching 
electric field, which is accompanied by the Rayleigh-type linear electric 
field dependence of permittivity due to the irreversible movement of the 
domain walls.
\cite{Taylor.JApplPhys.82.1997,Taylor.ApplPhysLett.73.1998,Xu.JApplPhys.89.2001,Gharb.JApplPhys.97.2005,Zhang.JApplPhys.100.2006,Gharb.JElectroceram.19.2007}
Nevertheless, one should note that there may arise experimental 
situations where even the practically small applied electric field breaks the 
limits for the irreversible movements of the domain walls and, therefore, 
violates the condition of the limit of the electric field tending zero. Thus, 
one should very carefully check that the experimental conditions satisfy the 
conditions for the applicability of the model.

The final advantage of the adopted method to study the nonlinear 
permittivity of the ferroelectric polydomain system in a weak electric field 
is that in this case, we do not need to strictly identify the microscopic 
origin of the domain wall pinning. The reason is that in weak electric fields 
the extrinsic contribution to the permittivity is controlled by fast 
reversible bending movements of the domain walls disregarding the nature of 
the pinning mechanism. Even if, in a particular sample, the domain wall 
pinning is strong due to many isolated crystal lattice impurities or it is 
weak due to the fluctuations of random fields in the vicinity of the domain 
wall, in a weak electric field, the pinning effect of both the random bonds or 
random fields results only in the bending of the domain wall. It means that 
our analysis can be reduced down to the problem of the identification of the 
bending movements of pinned domain walls. For that reason, the key 
elements of our analysis is the development of the model for the description 
of bending movements pinned 180${}^\circ$ domain walls. Within this model, we 
consider that the progressive pinning changes the bending condition of the 
domain wall during aging, which affects both the small signal permittivity 
$\varepsilon_L$ and the dielectric nonlinearity constant $b$.

We will show that if the dielectric aging is caused by progressive
pinning and is controlled by bending movements of 180$^{\circ}$ domain walls,
there exists the following relation between time dependencies of the
small-signal permittivity $\varepsilon_L(t)$ and the dielectric nonlinearity
constant $b(t)$:
\begin{equation}
  \sqrt{b(t)} \propto \varepsilon_L(t) - \varepsilon_c,
  \label{eq:02:bvseps}
\end{equation}
where $\varepsilon_c$ is the time-independent permittivity of the crystal 
lattice along the ferroelectric axis. To achieve the result announced above 
and given by Eqs.~(\ref{eq:01:EQuadr}) and (\ref{eq:02:bvseps}), we first 
present, in Sec. \ref{sec_model}, the details of our model of a ferroelectric 
film, where the movements of 180${}^\circ$ domain walls are locally blocked by 
pinning centers. By applying a straightforward thermodynamic methodology 
presented in Sec.~\ref{sec_thermodynamics}, we calculate the linear and 
nonlinear parts of the extrinsic contribution to permittivity controlled by 
bending movements of the pinned 180${}^\circ$ domain walls. In 
Sec.~\ref{sec_evolution}, we show that in the progressive pinning aging 
scenario there exists a characteristic relation between the linear and 
nonlinear domain wall contributions to permittivity of the polydomain 
ferroelectric. We demonstrate that by using measurements of the time evolution 
of the nonlinear permittivity it is possible to obtain evidence on whether the 
dielectric aging is controlled by the progressive pinning of 180${}^\circ$ 
domain walls or not. Section~\ref{sec_experiment} presents an application of 
our theory to aging experiments in [111]-oriented PZT (45/55) thin films.


\section{Model of 180${}^\circ$ domain wall bending}
\label{sec_model}

\begin{figure}[t]
	\includegraphics[width=70mm]{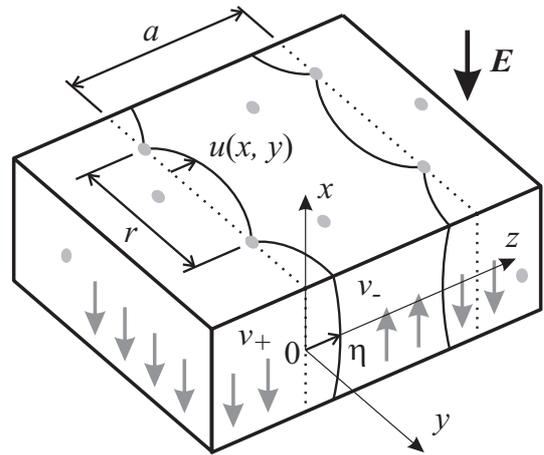}
\caption{
Model of 180${}^\circ$ domain wall bending. Planar position of the domain wall
(doted line) is locally blocked by pinning centers (gray circles). Orientation
of the vector of spontaneous polarization is indicated by gray arrows. When the
external electric field $E$ is applied to the ferroelectric film, the domain
wall is bent between pinning centers and the profile of the wall deflection is
described by the function $u(x,\,y)$.
}
\label{fig1}
\end{figure}

Figure \ref{fig1} shows the model
%
%
of a ferroelectric film with a lamellar 180${}^\circ$ ferroelectric domain
pattern, where the average distance between the domain walls (domain wall
spacing) is denoted by the symbol $a$. We consider that the top and bottom
electrodes of the film are perpendicular to the ferroelectric axis $x$ of the
attached Cartesian coordinate system. Using the ``hard ferroelectric''
approximation, we express the electric displacement within each ferroelectric
domain as a sum of the spontaneous polarization $P_{0}$ (whose orientation
differs from domain to domain) and the linear dielectric response of the
crystal lattice to the electric field,
\begin{subequations}
\label{eq:02:DIs}
\begin{eqnarray}
	\label{eq:02:DIsa}
	D_x &=&\varepsilon_0\varepsilon_c E_x \pm P_0 , \\
	\label{eq:02:DIsb}
   D_{y,z} &=&\varepsilon_0\varepsilon_a E_{y,z},
\end{eqnarray}
\end{subequations}
where $\varepsilon_{c}$ and $\varepsilon_{a}$ are the components of the
permittivity tensor of the crystal lattice in the directions along and
perpendicularly to the ferroelectric axis, respectively, and $\varepsilon_0$
is the permittivity of vacuum.

In the absence of the electric field, we consider the domain pattern to be
neutral, i.e., that the volume fractions $v_{+}$ and $v_{-}$ of the
adjacent antiparallel domains are the same, and that the 180${}^\circ$ domain
walls are parallel to the $x$-$y$ plane and pinned by immobile pinning
centers, which are distributed in the $x$ and $y$ directions with an average
distance $r$ between them. Thus, the pinning center density on the domain
wall is proportional to 1/$r^{2}$.

When the electric field $E$ is applied to the ferroelectric film, it exerts a
pressure $-2P_0E$ on the domain wall and it bends the originally planar domain
wall between the pinning centers. The domain wall bending produces a
change in the volume fractions $v_{+}$ and $v_{-}$ of the domains with the
spontaneous polarization oriented along and against the applied electric field,
 respectively. On the other hand, the domain wall deflection produces an
increase in the domain wall area. Since the bent domain wall is no longer
parallel to the vector of spontaneous polarization, a bound charge
$\sigma_{b}=-\Delta P_{0,n}$ appears on the domain wall due to the
discontinuous change in the normal component of spontaneous polarization at
the domain wall. If we denote the domain wall displacement in the
$z$ direction by the function $u(x,\,y)$, its exact profile is given by
minimizing the thermodynamic potential $G_w$, which consists of the
depolarizing field energy, the energy associated with the crystal lattice
polarization, the energy of the domain wall, and the energy supplied to the
system by the external electric source. In this model, it yields the solution
of Euler-Lagrange equation, which was analyzed earlier.
\cite{Yang.PhysRevLett.82.1999}

In order to calculate the extrinsic contribution to the permittivity due to 
bending movements of pinned domain walls, we approximate the generally random 
distribution of pinning centers by a periodic one with equidistant 
pinning centers, but with the same average density $1/r^2$. We consider that 
the pinning centers divide the infinite area of the domain wall into square 
segments with edges parallel to the $x$ and $y$ directions in such a way that 
each segment of the domain wall is pinned at its corners and the domain wall 
can move freely in the interior of each segment, as indicated in Fig. 
\ref{fig1}. This approximation reflects the assumption that the main 
contribution to the permittivity is coming from the structures on the domain 
wall with the periods that are close to the mean period in the system $r$. The 
strict procedure would be to expand the random distribution of the pinning 
centers in the Fourier series. Our approximation is good if the Fourier 
spectrum is not very wide. The consideration of the periodicity of pinning 
centers in the direction of the film thickness also limits the applicability 
of the domain wall bending model to samples of thicknesses much greater than 
the average pinning centers distance, i.e., $h\gg r$.

Finally, in most ferroelectric materials as well as in the samples used in our 
experiments, it is easy to show that even small domain wall displacements 
(about a lattice constant) are sufficient to produce the observed values of 
extrinsic contributions to the permittivity. Therefore, it is fully justified 
to consider that the maximum deflection $\eta$ of the domain wall is much 
smaller than the average distance between pinning centers $r$. In this case, 
if we take the origin of the coordinate system in the middle of each square 
segment, it is convenient to approximate the profile of the domain wall 
deflection in this segment by a parabolic function of the form,
\begin{equation}
	\label{eq:01:uG}
	u(x,\,y)=
		\eta
		\left[
			1-4\,\frac{\alpha\, x^2+y^2}{r^2(1+\alpha)}
		\right],
\end{equation}
where $\eta$ is the maximum deflection of the domain wall and the parameter
$\alpha$ is introduced in order to take into account the anisotropy of the
radius of curvature of the bent domain wall due to the strong depolarizing
effect in the direction of the ferroelectric $x$ axis. Coordinate values of
$x$ and $y$ are running over the interval from $-r/2$ to $r/2$.


\section{Extrinsic permittivity}
\label{sec_thermodynamics}

When the alternating electric field is applied to the ferroelectric sample,
the domain walls start to vibrate between the pinning centers, which results
in a change in the volume fractions $v_+$ and $v_-$ of the antiparallel
domains, where $v_++v_-=1$. This represents the source of the extrinsic
contribution to the permittivity of the ferroelectric film. To calculate the
extrinsic contribution to the permittivity, it is convenient to consider the
net spontaneous polarization $P_{N}$, which is given by the difference in the
volume fraction of domains with the vector of spontaneous polarization
oriented along and against the applied electric field, i.e.,
$P_N=\left(v_+-v_-\right)P_0$, and which can be expressed in the form,
\begin{equation}
	\label{eq:06:PNIs}
	P_N =\frac{2P_0 }{ar^2}\int_A {u(x,y)\;dA} =
	\frac{4\eta}{3a}\, P_0.
\end{equation}
where the integral is taken over the square segment $A$, while $x$ and $y$ are 
running from $-r/2$ to $r/2$. In the following text, we will use the above 
expression to measure the maximum deflection of the domain wall in terms of 
the net spontaneous polarization, i.e., $\eta = (3a\,P_N)/(4\,P_0)$. The 
response of the net spontaneous polarization with respect to the applied 
electric field $E$ is controlled by the thermodynamic function $G_w$ per unit 
volume of the ferroelectric,
\begin{equation}
	\label{eq:03:GwG}
 	G_w = \Phi_{s}(P_N) + \Phi_{\rm dep}(P_N) - P_N E,
\end{equation}
where the functions
\begin{equation}
	\label{eq:03:PhiS}
 	\Phi_{s} =
		\frac{1}{ar^2} \int_A
		\sigma _w 	\,	\sqrt {1+u_x^2 (x,\,y)+u_y^2 (x,\,y)} \;dA
\end{equation}
and
%
\begin{equation}
	\label{eq:03:PhiDep}
	\Phi_{\rm dep}
	=
	\frac{1}{ar^2}
	\int_A
		\frac 12 \sigma _b\, \varphi _d\,
	\sqrt {1+u_x^2 (x,y)+u_y^2 (x,y)} \;dA
\end{equation}
represent two contributions to the thermodynamic function $G_w$ due to the
increase in the domain wall area and due to the depolarizing field,
respectively. In Eqs.~(\ref{eq:03:PhiS}) and (\ref{eq:03:PhiDep}), the
integrals are taken over the square segment $A$ where $x$ and $y$ are running
from $-r/2$ to $r/2$, the symbol $\sigma_{w}$ stands for the surface energy
density associated with the surface tension of the bent domain wall, and the
symbol $\varphi_{d}$ stands for the electrostatic potential on the domain wall
associated with the depolarizing field, which is produced by the bound charges
$\sigma_{b}$.

The function $\Phi_{s}$ can be calculated in a straightforward way by direct
substitution of Eq.~(\ref{eq:01:uG}) into Eq.~(\ref{eq:03:PhiS}). The leading
terms of $\Phi_{s}$ with respect to the net spontaneous polarization equal
\begin{equation}
  \label{eq:09:PhiSIs}
	\Phi_{s} \approx
		\frac{
			3a \sigma_{w} \left(1+\alpha^2\right)
		}{
			4r^2\,P_0^2 \left(1+\alpha \right)^2
		} P_N^2
		-
  		\frac{
			9a^3 \sigma_{w} \left(9 + 10\alpha^2 + 9\alpha^4\right)
		}{
			80\,P_0^4\,r^4\,\left(1 + \alpha\right)^4
		} P_N^4.
\end{equation}
In order to express the function $\Phi_{\rm dep}$, it is necessary to 
calculate the spatial distribution of the electrostatic potential $\varphi_d$ 
in the vicinity of the bent domain wall. Since this goes beyond the main scope 
of this paper, the detailed calculations of the functions $\varphi_d$ and 
$\Phi_{\rm dep}$ are presented in the Appendix \ref{sec:Apendix}. Here we 
present only the result for the leading terms of the function $\Phi_{\rm dep}$ 
with respect to the net spontaneous polarization,
\begin{equation}
  \label{eq:10:PhiDepIs}
	\Phi_{\rm dep}
	\approx
	\frac{
		0.174 a\alpha^2
	}{
		\varepsilon_0r(1\!+\!\alpha)^2\sqrt{\varepsilon_a\varepsilon_c}
	}
	\left[
		P_N^2 -
      \frac{
			3a^2
			\left(
				1\!+\!1.43\alpha^2
			\right)
		}{
			2 P_0^2r^2(1\!+\!\alpha)^2
		}P_N^4
	\right].
\end{equation}

Now, the thermodynamic potential $G_w$ per unit volume of ferroelectric can
be expressed in the form of a Taylor expansion with respect to $P_{N}$. 
Disregarding the constant term, this function reads as
\ifthenelse{\equal{\twocolumnmode}{true}}{
%
%
%
%
\begin{eqnarray}
	\nonumber
	G_w
	&\approx&
	\frac{3a P_N^2}{4r^2(1\!+\!\alpha )^2}
	\left[
		\sigma _w
		\frac{(\alpha^2\!+\!1)}{P_0^2 }
		+
		\frac{7.2\,r\alpha ^2}{\pi ^3\varepsilon _0 \sqrt {\varepsilon _a \varepsilon _c } }
	\right]
  \\ \nonumber	
  &&-\frac{
    9 a^3 P_N^4
  }{
    80P_0^2 r^4(1\!+\!\alpha)^4
  }
  \left[
	 \sigma_w
    \frac{
      (9\!+\!10\alpha^2\!+\!9\alpha^4)
    }{
      P_0^2
    }
  \right.
  \\
  && +
	\left.
    \frac{r \alpha^2(2.33\!+\!3.32\alpha^2)
    }{
      \varepsilon_0\sqrt{\varepsilon_a\varepsilon_c}
    }
  \right]
  - P_N E,
	\label{eq:07:GwIs}
\end{eqnarray}
}{
%
%
\begin{eqnarray}
	\label{eq:07:GwIs}
	G_w
	&\approx&
	\frac{3a P_N^2}{4r^2(1\!+\!\alpha )^2}
	\left[
		\sigma _w
		\frac{(\alpha^2\!+\!1)}{P_0^2 }
		+
		\frac{7.2\,r\alpha ^2}{\pi ^3\varepsilon _0 \sqrt {\varepsilon _a \varepsilon _c } }
	\right]
	-
  \\ \nonumber
  \lefteqn{-
  \frac{
    9 a^3 P_N^4
  }{
    80P_0^2 r^4(1\!+\!\alpha)^4
  }
  \left[
	\sigma_w
	\frac{
      (9\!+\!10\alpha^2\!+\!9\alpha^4)
    }{
      P_0^2
    }
    +
    \frac{r \alpha^2(2.33\!+\!3.32\alpha^2)
    }{
      \varepsilon_0\sqrt{\varepsilon_a\varepsilon_c}
    }
  \right]
  - P_N E,
  }
\end{eqnarray}
}
where the first term in each square bracket is due to the increase in the
energy of the surface tension of the bent domain wall and the second term is
due to the increase in the depolarizing field energy produced by the bound
charges at the domain wall.

In what follows, it will be convenient to express the surface energy density
of the domain wall $\sigma_{w}$ in the form,
\begin{equation}
	\label{eq:08:Sigmaw}
	\sigma _w \approx \frac{a_w P_0^2 }{6\varepsilon _0 \varepsilon _c },
\end{equation}
where the parameter $a_{w}$ is of the order of the domain wall thickness. The
unknown value of the parameter $\alpha$ is determined from the condition
\begin{equation}
	\label{eq:09:AlphaCond}
	\partial G_w(P_N ,E;\alpha )/\partial \alpha =0.
\end{equation}
Later, it will be checked that in the case of a stable domain pattern and for
weak applied fields, the function $G_w$ is dominated by the lowest (quadratic)
term. Under this consideration, we can find the minimum of the function $G_w$ 
with respect to $\alpha$ and the condition given by Eq.~(\ref{eq:09:AlphaCond})
 yields
\begin{equation}
	\label{eq:10:AlphaMin}
	\alpha_{\min} = \left(1 + \frac{1.4\, r}{a_w}\sqrt{\frac{\varepsilon_c}{\varepsilon_a}}\right)^{-1}.
\end{equation}

In most of the samples, it is reasonable to consider that the domain wall
thickness $a_{w}$ is much smaller than the average distance between the pinning
centers $r$, i.e., $a_{w}\ll r$, and, in this case, Eq.~(\ref{eq:10:AlphaMin})
can be further simplified as
\begin{equation}
	\label{eq:11:AlphaMinA}
	\alpha _{\min } \approx
	\frac{a_w}{1.4\, r}\sqrt{\frac{\varepsilon_a}{\varepsilon_c}}.
\end{equation}
After substitution of Eqs. (\ref{eq:08:Sigmaw}) and (\ref{eq:11:AlphaMinA})
 into Eq. (\ref{eq:07:GwIs}), we obtain the following form for the expansion of
the function $G_w$ with respect to the net spontaneous polarization:
\begin{equation}
	\label{eq:12:GwTaylor}
	G_w(P_N,\,E)\approx
	\frac{a\,a_w }{8\varepsilon _0 \varepsilon _c r^2}P_N^2 -
	\frac{0.17\,a^3a_w }{\varepsilon _0 \varepsilon _c r^4 P_0^2 }P_N^4 - P_N E.
\end{equation}
It should be noted that in the case of most of the high-quality samples,
$\alpha_{min}\ll 1$ and the thermodynamic potential $G_w$ is dominated by the
surface tension, which can be readily seen after substituting Eqs.
(\ref{eq:08:Sigmaw}) and (\ref{eq:11:AlphaMinA}) into Eqs. (\ref{eq:09:PhiSIs})
 and (\ref{eq:10:PhiDepIs}).

The response of the net spontaneous polarization $P_{N}$ to the applied
electric field $E$ can be found from the condition for the minimum of the 
function $G_w$,
\begin{equation}
	\label{eq:13:GwCond}
	\partial G_w(P_N,\,E)/\partial P_N =0
\end{equation}
and it can be expressed as a Taylor expansion with respect to the
applied electric field,
\begin{equation}
	\label{eq:14:PNTaylor}
	P_N(E) \approx
	\frac{4\varepsilon _0 \varepsilon _c r^2}{a\,a_w}\,E +
	\frac{172.8\,r^4\varepsilon _0^3 \varepsilon _c^3 }{a\,a_w^3 \,P_0^2}\,E^3.
\end{equation}

Now we use the fact that the average electric displacement of the polydomain
film along the ferroelectric axis $D_f$ is given by the sum of the linear
dielectric response of the crystal lattice to the electric field and the net
spontaneous polarization, i.e., $D_f(E) = \varepsilon_0\varepsilon_c\, E +
P_N(E)$. 
Later we will check that the dielectric nonlinearity of the whole
system is dominated by the bending mechanism and, in this case, the field
dependence of permittivity of the ferroelectric polydomain film can be
expressed in the form
\begin{equation}
	\label{eq:15:Epsf}
	\varepsilon_f (E)\approx \varepsilon _c +\varepsilon_w + bE^2,
\end{equation}
where $\varepsilon_c$ is the intrinsic permittivity along the spontaneous
polarization and
\begin{eqnarray}
	\label{eq:16:ChiW}
	\varepsilon_w(r,\, a) &\approx& \frac{4\varepsilon _c r^2}{a\,a_w },
	\\
	\label{eq:17:b}
	b(r,\, a) &\approx& \left(\frac{1}{4}\right)
	\frac{518.4\,r^4\varepsilon_0^2 \varepsilon_c^3 }{a\, a_w^3 P_0^2}
\end{eqnarray}
are the coefficients of the small-signal linear and the quadratic terms of the
extrinsic contribution to permittivity with respect to the applied electric
field, respectively. Here it should be noted that Eq. (\ref{eq:14:PNTaylor}), 
is actually derived for the case of the dc field dependence of the net 
spontaneous polarization on the applied electric field. On the other hand, the 
experimental part of the paper addresses the ac field amplitude dependence of 
the average (mean) permittivity. Therefore, the relations given by Eqs. 
(\ref{eq:16:ChiW})-(\ref{eq:17:b}) are expressed for the ac field amplitude 
dependence of the average (mean) permittivity and the coefficient $b$ differs 
from the corresponding term in Eq. (\ref{eq:14:PNTaylor}) by a factor 1/4.


\section{Progressive pinning of domain walls and dielectric aging}
\label{sec_evolution}

Now it is seen that values of the small-signal extrinsic permittivity
$\varepsilon_w$ and of the dielectric nonlinearity coefficient $b$ are
controlled by the domain pattern configuration, i.e., by the pinning center
average distance $r$ and by the domain spacing $a$. It is natural to expect
that the parameters $r$ and $a$ can evolve with time, which may represent a
source of aging of the dielectric response and, therefore, a source of the
time evolution of the field dependence of permittivity $\varepsilon_f (E)$.
However, one should first identify the thermodynamic force, which can drive
the system to evolve with time and which can be responsible for the aging
process.

First let us focus on the possible role of the domain spacing $a$ in the
process of aging. In real systems, the domain pattern is usually controlled by
the prehistory of the sample so that the domain spacing may be essentially
different from its equilibrium value at the given temperature. It means that
there clearly exists a thermodynamic force that drives the system to reach the
state with the equilibrium domain spacing. This thermodynamic force actually
originates from the competition between the energy of the domain walls, the
electrostatic energy, and the surface energy at the interface between the
ferroelectric and the electrode. 
\cite{Kopal.Ferroelectrics.202.1997,Kopal.Ferroelectrics.223.1999,Bratkovsky.PhysRevLett.84.2000,Mokry.PhysRevB.70.2004} 
Nevertheless, frequent observations of rather stable but essentially 
nonequilibrium domain patterns
\cite{Odagawa.JpnJApplPhys.45.2006,Fujimoto.JpnJApplPhys.43.2004}
represent a clear indication that these energies are usually much weaker than
the energies involved in the pinning-depinning processes during the domain
wall movement under the application of sub-switching electric fields.
Therefore, it is not very reasonable to consider that the possible change in
the domain spacing could be responsible for the aging of the dielectric 
response.

The second scenario, which can be described within our domain wall bending 
model, is the \emph{progressive pinning of the domain walls}. Although it has 
been already mentioned in Sec.~\ref{sec_Intro} that one can apply the domain 
wall bending model on the weak-field nonlinear permittivity data disregarding 
the origin of pinning, one particular mechanism in perovskite ferroelectrics 
to mention here is pinning by the orientation of dipole 
defects.\cite{Robels.JApplPhys.73.1993} This model is based on the interaction 
of the domain walls with the dipoles formed by an acceptor ion (e.g., 
Ni${}^{2+}$, Fe${}^{2+}$, etc.) at the Ti${}^{4+}$ site and an oxygen vacancy 
in the surrounding oxygen octahedron. Since the free energy of the dipole 
defect in ferroelectrics depends on its orientation with respect to the vector 
of spontaneous polarization, the thermodynamic force---which drives the dipole 
defects to align with the spontaneous polarization vector in each domain in 
order to minimize the free energy of the whole system---causes an increase in 
the number of pinning centers that blocks the domain wall movement under a 
weak electric field. Thus, the increase in the number of such pinning centers 
naturally results in the decrease in the average distance $r$ between them.

Our model for bending movements of 180${}^{\circ}$ domain walls makes it
possible to identify such aging process using measurements of nonlinear
permittivity. We can express the average distance between pinning centers $r$
from Eq.~(\ref{eq:16:ChiW}) and substitute it into the formula for the
dielectric nonlinearity coefficient $b$ given by Eq.~(\ref{eq:17:b}). This
gives the relation between the parameters $b$ and $\varepsilon_w$ in the form,
\begin{equation}
	\label{eq:18:bchiPIN}
	b\approx \frac{8.1\,\varepsilon _0^2 \varepsilon _c}{P_0^2}\,
	\left(\frac{a}{a_w }\right)\,\varepsilon _w^2.
\end{equation}
If such a relation between $\varepsilon_w$ and $b$, which is typical for this
aging process and which is quite different from the prediction of, e.g., the
Landau theory, is identified in the aging measurements, it gives a reasonable
hint that the progressive pinning mechanism is responsible for the dielectric
aging of polydomain ferroelectrics. Therefore, the principal result of this
work is that by considering the domain wall bending mechanism and using the
analysis of the time evolution of the nonlinear dielectric response, we can
provide evidence that the dielectric aging is controlled by the progressive
pinning of the 180${}^\circ$ domain walls.

In addition, we can study the evolution of the pinning centers concentration
at the domain walls. By combining Eqs.~(\ref{eq:16:ChiW}) and
(\ref{eq:17:b}) and by eliminating the average domain spacing, one can
estimate the time dependence of the average pinning center distance according
to the following formula:
\begin{equation}
	\label{eq:19:rEvolPIN}
	r(t) =
		\frac{
			0.175\, a_w P_0
		}{
			\varepsilon_0 \varepsilon_c
		}
		\sqrt{
			\frac{b(t)}{\varepsilon_w(t)}
		}. 		
\end{equation}
%


\section{Experimental verification}
\label{sec_experiment}

In this section, we present a direct way of the experimental verification of 
the considered model for bending movements of the pinned domain walls to 
PZT ferroelectric films of tetragonal symmetry. Our model can be 
directly applicable to [001]-oriented PZT films. Unfortunately, 
there exists a reason, which makes the use of [001]-oriented PZT films not very 
practical to demonstrate the relevance of the domain wall bending model. 
Namely, the domain composition of $c$ and/or $a$ domains is difficult to 
control in such films. Because of the large dielectric anisotropy of PZT, this 
makes it difficult to control the lattice contribution to the permittivity. On 
the other hand, it appears to be convenient to use [111]-oriented films; since 
their lattice contribution is unique and independent on the domain pattern 
configuration, all 180${}^\circ$ domain walls have the same orientation with 
respect to the applied electric field and, thus, they contribute identically 
to the dielectric response. These features are important in applying our model 
to experimental data. Since the application of our model to [111]-oriented PZT 
films requires some modifications, we describe the application of 
the domain wall bending model to the both particular cases 
separately.

\subsection{Application to [001]-oriented PZT films}
\label{sec_001-films}

To prove that the progressive aging scenario is responsible
for the evolution of the dielectric response, we need to distinguish the
extrinsic $\varepsilon_w+bE^2$ from the intrinsic $\varepsilon_{c}$
contributions to the field dependence of the permittivity $\varepsilon_f(E)$.
Since it is natural to expect that the intrinsic contribution to the
permittivity $\varepsilon_{c}$ does not change in time, we can compare the
measured small-signal permittivity $\varepsilon_L$, given by the formula,
\begin{equation}
	\varepsilon_L = \varepsilon_f(0) = \varepsilon_c + \varepsilon_w,
	\label{eq:20:EpsL}
\end{equation}
with the dielectric nonlinearity constant $b$, which can be determined from
the experimental data using the following expression:
\begin{equation}
	b  = \left[\varepsilon_f(E) - \varepsilon_L\right]/E^2.
	\label{eq:21:bM}
\end{equation}
Then, with the use of Eq.~(\ref{eq:20:EpsL}), the relationship given in
Eq.~(\ref{eq:18:bchiPIN}) can be rewritten in the form,
\begin{equation}
	b\approx \frac{8.1\,\varepsilon _0^2 \varepsilon _c}{P_0^2}\,
	\left(\frac{a}{a_w }\right)\,\left(\varepsilon_L-\varepsilon_c\right)^2.
\end{equation}
Finally, taking the square root of $b$ from the above equation, we arrive at
the following relationship between $\varepsilon_L$ and $b$, which can be
cross checked experimentally:
\begin{equation}
	\label{eq:22:bFitEq}
	\sqrt b \approx K \varepsilon_L  - B,
\end{equation}
where
\begin{equation}
	\label{eq:23:KBIs}
	K =
	\sqrt {
		\frac{8.1\,\varepsilon _0^2 \varepsilon _c }{P_0^2}
		\left(\frac{a}{a_w } \right)
	}, \qquad
	B = K \varepsilon _c.
\end{equation}
Therefore the validity of Eq. (\ref{eq:18:bchiPIN}) can be demonstrated by a
linear relationship between the values of $\sqrt{b}$ and the small-signal
dielectric permittivity $\varepsilon_{L}$.

\subsection{Application to [111]-oriented PZT films}

In the case of the [111]-oriented PZT films, one cannot apply our model in a 
such a straightforward way as it was done in Sec.~\ref{sec_001-films}. There 
are two main reasons for that. First, the bending movements of pinned 
180$^\circ$ domain walls contribute to the value of the 
permittivity in the direction of the vector of spontaneous polarization. 
Second, perovskite ferroelectrics are materials with a rather large dielectric 
anisotropy in the directions parallel $\varepsilon_c$ and perpendicular 
$\varepsilon_a$ to the orientation of the vector of spontaneous polarization. 
In order to obtain reasonable numerical estimation of the microstructural 
parameters of the domain pattern, one should take the aforementioned point 
into account while interpreting the dielectric nonlinearity measurements. In 
the following text, we will denote all physical quantities, which are measured 
on the [111] oriented film, by a star superscript ``$^\star$.''

\begin{figure}[t]
	\includegraphics[width=85mm]{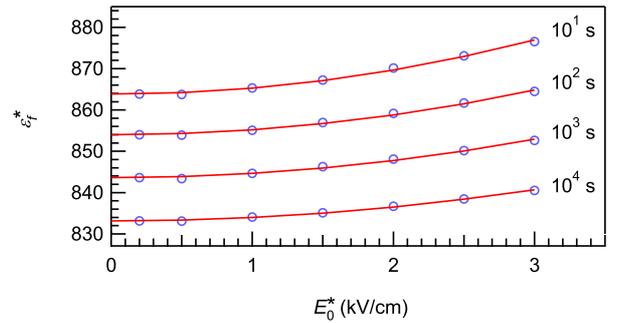}
	\caption{
(Color online) ac-field dependence of the dielectric permittivity (real part) 
within four decades of time. The measuring frequency was 1~kHz. The scattered 
markers represent the experimentally measured data. The solid lines are 
quadratic fittings using Eq. (\ref{eq:15:Epsf}).
}
\label{fig2}
\end{figure}
\begin{figure}[t]
	\includegraphics[width=85mm]{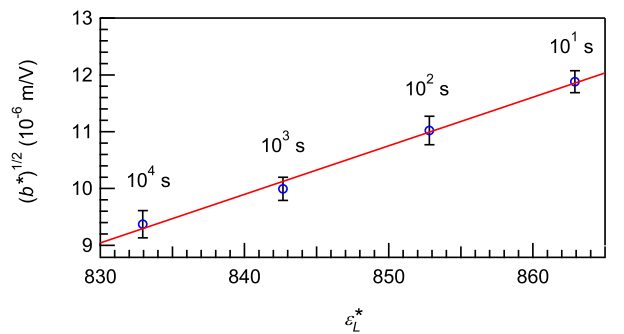}
\caption{
(Color online) Linear dependence of the square root of the coefficient 
$b^\star$ of the quadratic term in the expansion of the permittivity with 
respect to the applied electric field on the total weak-field permittivity of 
the ferroelectric film.
}
\label{fig3}
\end{figure}
The electric field dependence of the out-of-plane permittivity of the 
[111]-oriented PZT film in the tetragonal phase $\varepsilon_{f}^\star(E)$ is
given by the formula,
\begin{eqnarray}
	\varepsilon_{f}^\star(E)
	&=&
	\varepsilon_f(E)\,\cos^2(\theta)
	+ \varepsilon_a\,\sin^2(\theta )
	\\ \nonumber	
	&=&
	\left(
		\varepsilon_c + \varepsilon_w + b\,E^2		
	\right)\,\cos^2(\theta)
	+ \varepsilon_a\,\sin^2(\theta ),
\end{eqnarray}
where $E$ is the electric field along the vector of spontaneous polarization, 
the symbol $\theta$ stands for the angle between the spontaneous polarization 
and the normal to the plane of the film vector, which is equal to 
$\theta=\arctan(\sqrt{2})=54.7^\circ$. The value of $E$ is given by the 
applied electric field in the [111] direction $E^\star$, i.e., $E=E^\star 
\cos\theta$. If we introduce the dielectric anisotropy factor $\xi$ by the 
formula
\begin{equation}
	\xi = \varepsilon_a/\varepsilon_c,
	\label{eq:30:xiIs}
\end{equation}
the measured small-signal permittivity $\varepsilon_L^\star$ and the
dielectric nonlinearity constant $b^\star$ are then equal to
\begin{eqnarray}
	\label{eq:31:EpsLStarIs}
	\varepsilon_{L}^\star
	&=&
	\varepsilon_w\,{\cos}^2 (\theta )
	+
	\varepsilon_c\,
	\left[
		{\cos}^2 (\theta ) + \xi \,{\sin (\theta )}^2
	\right],
	\\
	\label{eq:31:bStarIs}
	b^\star	
	&=&
	b\,{\cos}^4 (\theta ).
\end{eqnarray}
Now we can follow the same procedure as in Sec.~\ref{sec_001-films}, i.e., we
express the value of $\varepsilon_w$ from Eq.~(\ref{eq:31:EpsLStarIs}) and we
combine it with Eqs.~(\ref{eq:18:bchiPIN}) and (\ref{eq:31:bStarIs}). Finally,
we obtain again the linear relationship between $\sqrt{b^\star}$ and
$\varepsilon_L^\star$,
\begin{equation}
	\label{eq:22:bStarFitEq}
	\sqrt{b^\star} \approx K^\star\, \varepsilon_L^\star  - B^\star,
\end{equation}
where
\begin{subequations}
\label{eq:23:KBStarIs}
\begin{eqnarray}
	\label{eq:23:KBStarIsK}
	K^\star
	&=&
	K,
	\\
	\label{eq:23:KBStarIsB}
	B^\star
	&=&
	\varepsilon _c	
	K\,
	\frac{
		\xi+1 - \left(\xi-1\right) \,{\cos} (2\theta)
	}2,
	\\
	\label{eq:23:KBStarIsC}
	K 
	&=&
	\sqrt {
		\frac{8.1\,\varepsilon _0^2 \varepsilon _c }{P_0^2}
		\left(\frac{a}{a_w } \right)
	}.
\end{eqnarray}
\end{subequations}


\section{Results and a discussion}
\label{sec_discussion}

The presented model of 180${}^\circ$ domain wall bending was applied to study 
the evolution of the dielectric response in a [111]-oriented 
Pb(Zr$_{0.45}$Ti$_{0.55})$O$_{3}$ thin film (240~nm in thickness). The film 
was deposited via a standard sol-gel method on a Pt-coated Si 
substrate.\cite{Taylor.ApplPhysLett.73.1998} Desired platinum patterns were 
vaporized on the surface of the crystallized film to form the top electrode. 
The out-of-plane dielectric response was measured using a Hewlett-Packard (HP)
~4284A high-precision impedance analyzer. The film was first depoled with a 
fast-decayed low-frequency (1~Hz) ac field (the amplitude decays from 80~kV/cm 
to zero in five periods). The dielectric response was then recorded as a 
function of the ac driving field $(E_{0}\le$3~kV/cm) and of the aging time at 
room temperature. The coercive field of the film is 60~kV/cm and, therefore, 
it is much larger than the maximum field used in this study.

Figure \ref{fig2} shows the ac-field dependence of the dielectric permittivity
within four decades of time, which can be well fitted by the quadratic relation
given in Eq. (\ref{eq:15:Epsf}). This is in contrast to the Rayleigh-type
relation, where the dielectric permittivity increases linearly with the
ac field as a result of the irreversible movement of domain walls under
subswitching fields (usually a few tens of kV/cm).
\cite{Taylor.JApplPhys.82.1997,Taylor.ApplPhysLett.73.1998,Xu.JApplPhys.89.2001,Gharb.JApplPhys.97.2005,Zhang.JApplPhys.100.2006,Gharb.JElectroceram.19.2007}
The bending of 180${}^\circ$ domain walls is actually a fast reversible
process. Figure \ref{fig3} shows the linear relationship between the values of
$\sqrt{b^\star}$ and the small-signal dielectric permittivity
$\varepsilon_{L}^\star$. Therefore, our experimental data unambiguously prove
the validity of Eq. (\ref{eq:22:bStarFitEq}).

Using the fit of results presented in Fig.~\ref{fig3}, we can estimate the 
lattice permittivity $\varepsilon_{c}$ and the ratio $a/a_{w}$. This is done 
in two steps. First, the slope $K^\star=85\times 10^{-9}\, {\rm mV^{-1}}$ and 
the offset $B^\star=62\times 10^{-6}\, {\rm mV^{-1}}$ of the linear dependence 
of $\sqrt{b^\star}$ versus $\varepsilon_L^\star$ are determined using a 
standard linear regression analysis. Second, the system of 
Eqs.~(\ref{eq:23:KBStarIs}) is solved by taking the values of spontaneous 
polarization $P_{0}=0.54~{\rm C/m}^{2}$ 
(Ref.~\cite{Amin.Ferroelectrics.65.1985}) and the dielectric anisotropy factor 
$\xi=3.3$.\cite{Haun.Ferroelectrics.99.1989} Solution of the system of 
Eqs.~(\ref{eq:23:KBStarIs}) gives a value for the lattice permittivity 
$\varepsilon_{c}$ equal to 280, which is in excellent agreement with the 
thermodynamic values,
\cite{Amin.Ferroelectrics.65.1985,Haun.Ferroelectrics.99.1989}
and for the ratio $a/a_{w}$ equal to $12\times 10^3$. From 
Eq.~(\ref{eq:31:EpsLStarIs}), the small-signal extrinsic contribution tothe 
permittivity along the spontaneous polarization $\varepsilon_w$ is estimated 
to be decreasing from 388 down to 296 during the time of 10$^{4}$~s of the 
aging experiment. The bending mechanism contributes to the total weak-field 
permittivity along the spontaneous polarization by about 50{\%}, which is also 
in agreement with the recent dielectric measurements of the weak-field 
permittivity of PZT ceramics in a very wide frequency 
range.\cite{Porokhonskyy.ToBePublished1} From the ratio $a/a_{w}=12\times 
10^3$. we can estimate the average domain spacing using the domain wall 
thickness as found from the Landau-Ginzburg-Devonshire theory (0.7~nm) 
(Ref.\cite{Hlinka.Ferroelectrics.375.2008}) or by an ab-initio calculation 
(0.5~nm).\cite{Meyer.PhysRevB.65.2002} The average distance between adjacent 
180${}^\circ$ domain walls thus has a value of about 5.9~${\rm \mu m}$. This 
specific distance is much larger than the typical grain size ($\sim $100~nm), 
indicating a low density of 180${}^\circ$ domain walls. This is reasonable in 
the case of strongly clamped films, as was shown by the transmission electron 
microscopy (TEM) observation in Ref.\cite{Xu.JApplPhys.89.2001} In addition, 
the average distance between adjacent 180${}^\circ$ domain walls is controlled 
by the prehistory of the sample and by the depoling process used before the 
aging experiment. Finally, our experiments indicate that the average distance 
between the pinning centers has decreased from 33 down to 29~nm during 
10$^{4}$~s. These values, which are much smaller than the film thickness, 
correspond to the volume concentration of crystal lattice impurities being 
higher than 0.03{\%}, which is acceptable since it is known that the 
nominally pure PZT ceramic films possess naturally occurring acceptor 
impurities.

Considering the key element of our method---the model of bending movements of 
pinned 180${}^\circ$ domain walls---we have calculated the linear and 
nonlinear contributions to extrinsic permittivity in the polydomain 
ferroelectric and used this result to analyze our experimental data. As a 
result, we were able to extract some microstructural parameters of the domain 
pattern, i.e., the average distance between the pinning centers and the domain 
spacing, and material parameters of the ferroelectric, i.e., the intrinsic 
permittivity of the crystal lattice. However, three important assumptions have 
been made during the development of the theoretical model and now it should be 
checked whether the values of the parameters fitted from our experimental data 
do not violate such assumptions: (a) the nonlinearity is dominated by bending 
of the 180${}^\circ$ domain walls, (b) the maximum deflection of the bent 
domain wall is much smaller than the average distance between the pinning 
centers, and (c) the free energy $G_w$ that controls the net spontaneous 
polarization response to the electric field $P_N(E)$ is dominated by the 
lowest (quadratic) term.

The first assumption can be verified by comparing the dielectric nonlinearity
constant that is controlled by bending movements of the 180${}^\circ$ domain
walls $b$ with the dielectric nonlinearity of the crystal lattice $b_c$,
which is equal to\cite{Mokry.Ferroelectrics.375.2008}
\begin{displaymath}
	b_c = \frac{12 \, \varepsilon_0^2\varepsilon_c^3}{P_0^2}
		\left(
			1 - \frac{3\varepsilon_w}{8\varepsilon_c}
		\right).
\end{displaymath}
By substituting the numerical results of our study, one can immediately see 
that $b_c=1.6\times 10^{-16}\,{\rm m^2V^{-2}}$, which is about 6 orders of 
magnitude smaller than the values observed in the experimental part of this 
study, i.e., $b\approx\, 8\times 10^{-10}\,{\rm m^2V^{-2}}$. 

The second assumption can be verified by using 
Eqs.~(\ref{eq:06:PNIs}) and (\ref{eq:14:PNTaylor}). After the substitution of 
the numerical parameters fitted from our experiments, we can obtain that, at 
the maximum fields applied to our sample, i.e., 3\,kV/cm, the maximum 
deflection $\eta$ of the domain wall is about 17~nm. This value is more than 
two times smaller than the estimated average distance between the pinning 
centers $r$ (over 40\,nm).

The third assumption can be verified by the substitution of
Eq.~(\ref{eq:06:PNIs}) into Eq.~(\ref{eq:07:GwIs}). With use of the numerical
parameters fitted from our experiments, we can see that, at the maximum fields
applied to our sample, i.e., 3\,kV/cm, the value of the lowest (quadratic) term
in the expansion of the thermodynamic function $G_w$ is about 302\,
J\,m${}^{-3}$. This value is more than three times greater than the value
of the higher- (fourth-) order term, which is approximately -90\,J\,m${}^{-3}$.

Therefore, it is seen that all assumptions made in the theoretical part of our
study are fully justified and applicable to real ferroelectric samples. By
applying the main results of the theoretical model [Eqs. (\ref{eq:15:Epsf}) to
(\ref{eq:18:bchiPIN})] to our experimental data, we have provided a strong
evidence that the dielectric nonlinearity in polydomain ferroelectrics is
predominantly controlled by the considered bending mechanism. In addition, we
have shown that from the evolution of the relation between the small-signal
permittivity $\varepsilon_L$ and the dielectric nonlinearity constant $b$
it is possible to distinguish the microstructural mechanisms, which are
responsible for the aging of the dielectric response. In our particular
experimental case, the linear relationship between the values of
$\sqrt{b^\star}$ and $\varepsilon_{L}^\star$ provided us with evidence that
the decrease in the linear dielectric permittivity during aging can be
attributed to the increase in the average pinning center density on the domain
wall, indicating a progressive pinning nature of the aging phenomenon. Finally,
 Eqs.~(\ref{eq:16:ChiW}) and (\ref{eq:17:b}) show that---from known values of
the parameters $\varepsilon_{L}$ and $b$ at a given time---it is possible to
estimate the actual values of the domain spacing $a$ and the average pinning
center density on the domain wall $1/r^{2}$. Therefore, we believe that our
results can be used as a simple and useful tool for getting a deeper insight
into the configuration of the domain pattern and the quality of ferroelectric
thin films and for providing a way to identify the evolution of the 
180${}^\circ$ domain pattern microstructure in perovskite 
ferroelectrics.


\acknowledgments

This work has been supported by the Czech Science Foundation under Project No. 
GACR~202/06/0411 and No. GACR~202/07/1289, by the Swiss National Science 
Foundation, and by the MIND - European network on Piezoelectrics. Authors 
thank Guido Gerra for reading the manuscript.

\appendix


\section{Electrostatic energy of the bent 180${}^\circ$ domain wall}
\label{sec:Apendix}

In this appendix, we present the detailed calculation of the electrostatic
potential $\varphi_d$ and the depolarizing field energy per unite
volume of the ferroelectric $\Phi_{\rm dep}$ with the bent 180${}^\circ$ domain 
walls. When the domain wall is bent, the discontinuous change in 
the normal component of spontaneous polarization $P_{0,n}$ at the domain wall 
yields the appearance of a bound charge of surface density $\sigma_{b}=-\Delta 
P_{0,n}$,
\begin{equation}
	\sigma_{b}(x,\,y) =
		\frac{
			-2P_0\, u_x(x,\,y)
		}{
			\sqrt {1+u_x^2(x,\,y)+u_y^2(x,\,y)},
		}	
	\label{eq:01:sigmab_gen}
\end{equation}
where the functions $u_{x}(x,y)$ and $u_{y}(x,y)$ are the partial derivatives 
of the function $u(x,y)$ with respect to $x$ and $y$, respectively. If the 
maximum deflection $\eta$ of the domain wall is much smaller than the average 
distance between pinning centers $r$ then by using 
Eq.~(\ref{eq:01:uG}), the bound charge density on the domain wall can be 
expressed as a Taylor expansion with respect to $\eta$ as follows:
\begin{equation}
	\sigma_{b}(x,\,y) \approx
		\frac{
			16\,P_0\, \alpha x
		}{
			r^2\left(1+\alpha\right)
		}\, \eta
		-
		\frac{
			512\,P_0\, \alpha x \left(y^2+\alpha^2 x^2\right)
		}{
			r^6\left(1+\alpha\right)^3
		}\, \eta^3.
	\label{eq:01:sigmab_ser}
\end{equation}

If we consider that the maximum deflection of the domain wall is much smaller 
than the average distance between the pinning centers, i.e., $\eta\ll r$, 
the electrostatic potential $\varphi_{d}$ can be approximated by 
the solution of the electrostatic problem where we assume that the bound 
charges $\sigma_{b}$ on the bent domain wall are located at the original 
position of the domain wall, i.e., at $z=0$. Then, the electrostatic potential 
$\varphi_{d}$ is given by the solution of the Laplace equation,
\begin{subequations}
\label{eq:04:Pot}
\begin{equation}
    \varepsilon_c\frac{\partial^2\varphi_d}{\partial x^2} +
    \varepsilon_a
	 \left(
		\frac{\partial^2\varphi_d}{\partial y^2} +
		\frac{\partial^2\varphi_d}{\partial z^2}
	\right)= 0,
	\label{eq:04:PotLap}
\end{equation}
%
%
%
with the internal boundary conditions for the continuity of the normal
component of electric displacement and for the continuity of the electrostatic
potential at the domain wall, i.e., at $z=0$,
\begin{equation}
   \frac{\partial\varphi_d^{(+)}}{\partial z}
	-
	\frac{\partial\varphi_d^{(-)}}{\partial z}
	=
   \frac{\sigma_b}{\varepsilon_0\varepsilon_a},
   \quad
   \varphi_d^{(+)}
	=
   \varphi_d^{(-)},
	\label{eq:04:PotBoundCond}
\end{equation}
\end{subequations}
where the superscripts $(+)$ and $(-)$ denote the electrostatic potential for
$z>0$ and $z<0$, respectively. Considering the periodicity of the bound charge
surface density on the domain wall in both the $x$ and $y$ directions with
the period equal to $r$, we will look for a solution of the system of
Eq.~(\ref{eq:04:Pot}) in the form of a Fourier series. It can be easily shown
that the functions
\ifthenelse{\equal{\twocolumnmode}{true}}{
%
%
\begin{eqnarray}
	\nonumber
	\varphi_d^{(\pm)}(x,\,y,\,z)
	&=&
	\sum_{n=1}^\infty
	\phi_{n0}
	\sin \left(\frac{2n\pi x}{r}\right)
	e^{
		\mp(2n\pi z/r)\,
		\sqrt{\varepsilon_c/\varepsilon_a}
	} 
	\\ \nonumber \lefteqn{
	+ \sum_{n,m=1}^\infty
	\phi_{nm}
	\sin \left(\frac{2n\pi x}{r}\right)\cos \left(\frac{2m\pi y}{r}\right)
	}
	\\ \label{eq:05:PotFourierGen}
	\lefteqn{
	\times
	e^{
		\mp(2\pi z/r)\,
		\sqrt{m^2 + n^2\,\varepsilon_c/\varepsilon_a}
	}
	}
\end{eqnarray}
}{
%
%
%
\begin{eqnarray}
	\label{eq:05:PotFourierGen}
	\varphi_d^{(\pm)}(x,\,y,\,z)
	&=&
	\sum_{n=1}^\infty
	\phi_{n0}
	\sin \left(\frac{2n\pi x}{r}\right)
	e^{
		\mp(2n\pi z/r)\,
		\sqrt{\varepsilon_c/\varepsilon_a}
	} +
	\\ \nonumber
	\lefteqn{
	+
	\sum_{n,m=1}^\infty
	\phi_{nm}
	\sin \left(\frac{2n\pi x}{r}\right)\cos \left(\frac{2m\pi y}{r}\right)
	e^{
		\mp(2\pi z/r)\,
		\sqrt{m^2 + n^2\,\varepsilon_c/\varepsilon_a}
	}
	}
\end{eqnarray}
}
satisfy the Laplace (\ref{eq:04:PotLap}). The unknown coefficients
$\phi_{n0}$ and $\phi_{nm}$ can be found by substituting
Eqs.~(\ref{eq:01:sigmab_ser}) and (\ref{eq:05:PotFourierGen}) into
Eq.~(\ref{eq:04:PotBoundCond}). Straightforward calculations yield the
formulae for the unknown coefficients,
\ifthenelse{\equal{\twocolumnmode}{true}}{
%
%
\begin{subequations}
\label{eq:06:PhiCoefs}
\begin{eqnarray}
	\label{eq:06:PhiN0CoefIs}
	\phi_{n0}
	&=&
	-\frac{
		3a\alpha\,(-1)^n
	}{
		n^2 \pi^2\varepsilon_0 \sqrt{\varepsilon_a\varepsilon_c}(1+\alpha)
	}
	\times \\ \nonumber \lefteqn{
	\left\{
		 P_N
		+
		\frac{
			3a^2\, \left[18 \alpha^2 - n^2\pi^2\, (1 + 3\alpha^2)\right]
		}{
			2n^2\pi^2r^2P_0^2 (1+\alpha)^2
		}\, P_N^3
	\right\},
	}
	\\
	\phi_{nm}
	&=&
	\frac{
		54a^3\alpha\,(-1)^{m+n}
	}{
		m^2 n \pi^4 \varepsilon_0 r^2 P_0^2
		\sqrt{m^2\,\varepsilon_a^2 + n^2\,\varepsilon_c\varepsilon_a}
		(1+\alpha)^3\,    	
	}\,P_N^3.
	\nonumber \\[-1ex]
	\label{eq:06:PhiNMCoefIs}
\end{eqnarray}
\end{subequations}
}{
%
%
\begin{subequations}
\label{eq:06:PhiCoefs}
\begin{eqnarray}
	\label{eq:06:PhiN0CoefIs}
	\phi_{n0}
	&=&
	-\frac{
		3a\alpha\,(-1)^n
	}{
		n^2 \pi^2\varepsilon_0 \sqrt{\varepsilon_a\varepsilon_c}(1+\alpha)
	}
	\left\{
		 P_N
		+
		\frac{
			3a^2\, \left[18 \alpha^2 - n^2\pi^2\, (1 + 3\alpha^2)\right]
		}{
			2n^2\pi^2r^2P_0^2 (1+\alpha)^2
		}\, P_N^3
	\right\},
	\\
	\phi_{nm}
	&=&
	\frac{
		54a^3\alpha\,(-1)^{m+n}
	}{
		m^2 n \pi^4 \varepsilon_0 r^2 P_0^2
		\sqrt{m^2\,\varepsilon_a^2 + n^2\,\varepsilon_c\varepsilon_a}
		(1+\alpha)^3\,    	
	}\,P_N^3.
	\label{eq:06:PhiNMCoefIs}
\end{eqnarray}
\end{subequations}
}
After substitution of Eq.~(\ref{eq:01:sigmab_gen}) into
Eq.~(\ref{eq:03:PhiDep}), the function $\Phi_{\rm dep}$ can be expressed in the
following forms:
\ifthenelse{\equal{\twocolumnmode}{true}}{
%
%
\begin{eqnarray}
	\label{eq:07:PhiDepSubstSum}
	\Phi_{\rm dep}
	&=&
	-\frac{P_0}{ar^2}
	\int_A u_x(x,y)\, \varphi _d(x,\,y,\,0)\;dA
	\\ \nonumber
	&=&
	\sum_{n=1}^\infty
	\frac{-6\alpha\,(-1)^n\,P_N}{n\pi r\,(1+\alpha)}\,\phi_{n0}.
\end{eqnarray}
}{
%
%
\begin{equation}
	\label{eq:07:PhiDepSubstSum}
	\Phi_{\rm dep}
	=
	-\frac{P_0}{ar^2}
	\int_A u_x(x,y)\, \varphi _d(x,\,y,\,0)\;dA
	=
	\sum_{n=1}^\infty
	\frac{-6\alpha\,(-1)^n\,P_N}{n\pi r\,(1+\alpha)}\,\phi_{n0}.
\end{equation}
}
It should be noted that the coefficients $\phi_{nm}$ do not enter the formula 
for $\Phi_{\rm dep}$ because the function $u_x(x,y)$ does not depend on $y$ 
and the integration of all terms with $\cos(2m\pi y/r)$ over its period $r$ 
gives zero. After further substitution of Eq.~(\ref{eq:06:PhiCoefs}) into 
Eq.~(\ref{eq:07:PhiDepSubstSum}) and performing the summation, the Taylor 
expansion of the $\Phi_{\rm dep}$ with respect to the net spontaneous 
polarization is equal to
\begin{equation}
	\Phi_{\rm dep}
	\approx
	\frac{
		0.174 a\alpha^2
	}{
		\varepsilon_0r(1\!+\!\alpha)^2\sqrt{\varepsilon_a\varepsilon_c}
	}
	\,\left[
		P_N^2\! -\!
      \frac{
			3a^2
			\left(
				1\!+\!1.43\alpha^2
			\right)
		}{
			2 P_0^2r^2(1\!+\!\alpha)^2
		}P_N^4
	\right].
\end{equation}


\bibliographystyle{apsrev}
\bibliography{mokry_cond-mat_v4}

\begin{thebibliography}{28}
\bibitem[{\citenamefont{Porokhonskyy and
  et.al.}()}]{Porokhonskyy.ToBePublished1}
\bibinfo{author}{\bibfnamefont{V.}~\bibnamefont{Porokhonskyy}}
  \bibnamefont{and} \bibinfo{author}{\bibnamefont{et.al.}}, \bibinfo{note}{(to
  be published)}.
\end{thebibliography}

%

\end{document}